\magnification=1250 \vsize=25truecm \hsize=16truecm
\baselineskip=0.6truecm \parindent=0truecm \parskip=0.2cm \hfuzz=0.8truecm

\font\scap=cmcsc10

\font\tenmsb=msbm10
\font\sevenmsb=msbm7
\font\fivemsb=msbm5
\newfam\msbfam
\textfont\msbfam=\tenmsb
\scriptfont\msbfam=\sevenmsb
\scriptscriptfont\msbfam=\fivemsb

\nopagenumbers

\def\xup{\overline x}
\def\Xup{\overline  X}

\def\yup{\overline  y}

\def\zup{\overline  z}

\def\xdo{\underline x}

\null\bigskip\centerline{\bf Integrable systems without the
Painlev\'e property}
\vskip 2truecm
\bigskip
\centerline{\scap A. Ramani}
\centerline{\sl CPT, Ecole Polytechnique}
\centerline{\sl CNRS, UMR 7644}
\centerline{\sl 91128 Palaiseau, France}
\bigskip
\centerline{\scap B. Grammaticos}
\centerline{\sl GMPIB, Universit\'e Paris VII}
\centerline{\sl Tour 24-14, 5$^e$ \'etage, case 7021}
\centerline{\sl 75251 Paris, France}\bigskip
\centerline{\scap S. Tremblay}
\centerline{\sl Centre de Recherches Math\'ematiques}
\centerline{\sl et D\'epartement de Physique}
\centerline{\sl Universit\'e de Montr\'eal}
\centerline{\sl C.P. 6128, Succ.~Centre-ville}
\centerline{\sl Montr\'eal, H3C 3J7, Canada}\bigskip\bigskip
Abstract
\medskip
We examine whether the  Painlev\'e property is a necessary condition for
the integrability of nonlinear ordinary
differential equations. We show that for a large class of linearisable
systems this is not the case. In the
discrete domain, we investigate whether the singularity confinement
property is satisfied for the discrete
analogues of the non-Painlev\'e continuous linearisable systems. We find
that while these discrete
systems are themselves linearisable, they possess nonconfined singularities.
\vfill\eject
\footline={\hfill\folio} \pageno=2
\noindent {\scap 1. Introduction }
\medskip
For over a century, the Painlev\'e property [1] has been  the cornerstone
of integrability. The reason Painlev\'e
introduced this property, which later was called after him, was a question
that was open
at the time, and of particular interest: `Is it possible to define (new)
functions from the solutions of
nonlinear differential equations?'

In some sense, this amounted to introducing the analogue of special
functions into the nonlinear domain. The
study of linear equations had shown where the difficulties lied [2]. In
particular, one had to deal with the
multivaluedness that could appear as a consequence of the singularities of
the coefficients of the equation which,
for  linear equations, are the only possible singularities of the
solutions. The extension of these
ideas to the nonlinear domain appeared hopeless since the location of bad
singularities could now depend on
the initial conditions. Then Painlev\'e made a leap of faith by requesting
that all critical (i.e. multivalued)
movable (i.e. initial condition dependent) singularities be absent.

The ordinary differential equations (ODE's) without critical  movable
singularities are said to possess the
Painlev\'e property. Their solutions define functions which in some cases
(the Painlev\'e transcendents) cannot be
expressed in terms of known functions. The precise way to integrate (i.e.
to construct the solutions of) the
ODE's with the Painlev\'e property can be very complicated [3] but the
important fact is that this can in principle
be done. Thus, the property came to be synonymous to integrability.
At this point it must be made clear that the integrability we are talking
about,  related to the Painlev\'e
property,  is of a special kind often referred to as `algebraic
integrability' [4]. It
is, for instance, the kind of integrability that characterises systems
integrable in terms of Inverse
Scattering Transform (IST) techniques [5]. However, in common practice,
many other `brands' of  integrability do
exist [6]. Integrability through quadratures, like that encountered in the
case of finite-dimensional Hamiltonian
systems, is of (relatively) frequent occurence, and is not identical to
algebraic integrability.
Linearisability, i.e. the reduction of the system to a system of linear
equations through  local
transformation, is a further, different, type.

In this paper, we shall examine the relation of these kinds of
integrability to the Painlev\'e property,
focusing on
linearisable systems. In the second part of the paper, we shall examine the
discrete analogues of these
notions. In this case, the role of the Painlev\'e property is played by
singularity confinement [7]. The
latter is believed to be a necessary condition for integrability (but
unlike the Painlev\'e
property it has turned out not to be sufficient as well [8]). We shall show
that in both continuous and
discrete settings, linearisable systems integrable through linearisation
can exist  without the
Painlev\'e property.
\bigskip
{\scap 2. Integrable continuous systems and the Painlev\'e property.}
\smallskip
A first instance of integrability without the Painlev\'e property was the
derivation of the integrable
system described by the Hamiltonian [9]:
$$H= {1\over 2} p_x^2+{1\over 2}p_y^2 +y^5+y^3x^2+{3\over 16}yx^4\eqno(2.1)$$
which has the second (besides the energy) constant of motion
$$C= -yp_x^2+xp_xp_y +{1\over 2}y^4x^2+{3\over 8}y^2x^4+{1\over
32}x^6.\eqno(2.2)$$
There are movable singularities where near some singular point $t_0$, one
has $y\approx \alpha(t-t_0)^{-2/3}$,
$x\approx \beta(t-t_0)^{-1/3}$ with $\alpha^3=-2/9$, $\beta$ arbitrary.
Taking the cube of the variables is
not sufficient to regularise them, however. Indeed, a detailed analysis of
complex-time singularities
shows that their expansions  contain {\sl all} powers of
$(t-t_0)^{1/3}$.  The fact that some multivaluedness was compatible with
integrability led to the introduction of the notion of ``weak Painlev\'e''
property. However, it was soon
realised [10] that (2.1) was a member of a vaster family of  integrable
Hamiltonian systems associated
to the potential
$V=(F(\rho+y)+G(\rho-y))/\rho$ where $\rho=\sqrt{x^2+y^2}$. Since the two
functions $F$ and $G$ are free, one
can easily show that the singularities of the solutions of the equations of
motion can be  arbitrary.
The Hamiltonians of this family are integrable through quadratures and, in
fact, the associated Hamilton-Jacobi
equations are separable.
This leads to the conclusion that this type of
integrability is not necessarily related to the  Painlev\'e property. (As a
matter of fact, the same conclusion
could have been reached if we had simply considered one-dimensional
Hamiltonian systems).
One may justifiably argue that in the case of Hamiltonian systems the term
integrability is to be understood as
Liouville integrability which is not the one we refer to in relation to the
Painlev\'e property. Still, Liouville
integrability, and the dynamical symmetries to which it is associated, may
be of utmost importance for physical
applications and a systematic method for the detection would have been most
welcome.

We turn now to a second case of
integrability where the necessary character of the Painlev\'e property can
be critically examined: that
of linearisable systems. The term  linearisable is used here to denote
systems that can be reduced to linear
equations through a local variable transformation. The first family of such
systems are the projective
ones [11]. Starting from the linear system for ($N$+1) variables:
$$X'_\mu=\sum_{\nu=0}^N A_{\mu\nu}X_\nu \qquad\mu=0,1,\dots, N \eqno(2.3)$$
and introducing the quantities $x_\mu=X_\mu/X_0$ we obtain the projective
Riccati system:
$$x'_\mu=a_\mu+\sum_{\nu=1}^N b_{\mu\nu}x_\nu+x_\mu\sum_{\nu=1}^N
c_{\mu\nu}x_\nu \qquad\mu=1,\dots, N
\eqno(2.4)$$
where $a_\mu$, $b_{\mu\nu}$ and $c_{\mu\nu}$ are given in terms of
$A_{\mu\nu}$.
As we have shown in [12] this system can be rewritten as a single $N$-th
order differential equation.
For $N$=1, this is just the Riccati equation for $x_1$:
$$x'_1=a_1+b_{11} x_1+c_{11}x_1^2 \eqno(2.5)$$
For $N$=2, the system  can be reduced to the equation VI of the
Painlev\'e/Gambier classification [2]
$${d^2w\over dz^2}=-3w{dw\over dz}-w^3+q(z)({dw\over dz}+w^2) \eqno(2.6)$$
for $z$ some function of the independent variable of (2.4) and $w$ a
homographic function of $x_1$ with
some specific functions of $z$ as coefficients.
Because of the underlying linearisation, the projective Riccati systems
possess the Painlev\'e property by
construction. Indeed, the $X_\mu$ have no movable singularities at all, and
the only movable singularities of the
$x_\mu$ are poles coming from the zeros of $X_0$.

However, there exists another kind of linearisability for which the
Painlev\'e property need not be satisfied.
Let us discuss the best-known second order case. One of the equations of
the Painlev\'e/Gambier classification,
bearing the number XXVII, is the equation proposed by Gambier [13]:
$$x''={n-1\over n}{x'^2\over x}+\left(fx+\phi-{n-2\over
nx}\right)x'-{nf^2\over (n+2)^2}x^3
+{n(f'-f\phi)\over n+2}x^2+\psi x-\phi-{1\over nx}\eqno(2.7)$$
where $f$, $\phi$ and $\psi$ are definite rational functions of two
arbitrary analytic functions and of their
derivatives [2]. As Gambier has shown, equation (2.7)
can be written as a system of two Riccati equations in cascade:
$$y'=-y^2+\phi y+{2f\over n(n+2)}+{\psi\over n}\eqno(2.8a)$$
$$x'={nf\over n+2}x^2+nyx+1.\eqno(2.8b)$$
Gambier has shown that unless the parameter $n$ appearing in (2.7) and
(2.8) is integer, the equation does
not possess the Painlev\'e property. (We must point out that this is a
first necessary condition and, in
general, not a sufficient one: further constraints on the coefficients are
needed in order to
ensure the Painlev\'e property).
On the other hand, the integration of the two Riccati equations in cascade
can always be performed, through
reduction to linear second order equations. Thus, although the solution of
(2.8) does not in general lead to a
well-defined function as solution of (2.7), it can still be obtained in
cascade.

Once the Painlev\'e property is deemed unnecessary for the linearisation of
the Gambier system, it is
straightforward to extend the latter to the form:
$$y'=\alpha y^2+\beta y+\gamma\eqno(2.9a)$$
$$x'=a(y,t)x^2+b(y,t)x+c(y,t)\eqno(2.9b)$$
where $\alpha$, $\beta$ and $\gamma$ are arbitrary functions of  $t$ while
$a$, $b$ and $c$ are arbitrary
functions of $y$ and $t$. The integration in cascade of (2.9) can be
obtained as previously. As a matter of fact,
an extension like (2.9) gives the handle to the ($N$+1)-variables
generalisation of the Gambier system:
$$x_0'=a_0(t)x_0^2+b_0(t)x_0+c_0(t)\eqno(2.10)$$
$$x_\mu'=a_\mu(x_0,\dots,x_{\mu-1},t)x_\mu^2+b_\mu(x_0,\dots,x_{\mu-1},t)x_\mu+c
_\mu(x_0,\dots,x_{\mu-1},t)
\qquad\mu=1,\dots, N$$
where $a_\mu$, $b_\mu$ and $c_\mu$ are arbitrary functions of their arguments.
Again, system (2.10) does not possess, generically, the Painlev\'e property
while it can be linearised and
integrated in cascade.

Untill now, we have presented rather straightforward generalisations of
integrable systems which violate the
Painlev\'e property while preserving their linearisability. We shall close
this section by introducing a new
(at least to our knowledge) method of linearisation which again leads to
integrable systems not
possessing the Painlev\'e property. Let us describe our general approach.
The idea is the following: we start
from a linear second order equation in the form:
$${\alpha x''+\beta x'+\gamma x+\delta\over \epsilon x''+\zeta x'+\eta
x+\theta}=K\eqno(2.11)$$
where $\alpha,\beta,\dots, \theta$ are functions of $t$ with $K$ a
constant, and a {\sl nonlinear} second
order equation of the form:
$$f(x'',x',x)=M\eqno(2.12)$$
where $f$ is  a (possibly inhomogeneous) polynomial of
degree  two in
$x$ together with its derivatives, but linear in $x''$, and with $M$ a
constant. We then ask that the
derivatives of both equations with respect to the independent variable,
i.e. the resulting third order
equations, be
identical up to an overall factor.
This is a novel linearisation approach.  The explicit integration procedure
is the following. We start from
equation (2.12) with {\sl given} $M$ and initial conditions $x_0$, $x'_0$
for some value $t_0$ of the
independent  variable $t$. We use (2.12) to compute $x''_0$ at  $t_0$.
Having these values, we
can use (2.11) to compute the value of $K$. Since the latter is assumed to
be a constant, we can
integrate the linear equation (2.11) for all values of $t$.  Since this
solution will
satisfy the third order equation mentioned above, it will also be a
solution of (2.12).

In order to illustrate this approach, we derive one equation that can be
integrated through this linearisation.
Our starting assumption is that (2.12) contains a term $x''x'$. The more
general term $x''(x'+cx+d)$ can always be
reduced to this form, i.e. $c$=$d$=$0$ through a rescaling and translation
of $x$. It is then straightforward
to obtain the full expression in the homogeneous subcase $\delta=\theta=0$.
We thus find:
$${t x''+(at-1/2) x'+bt x\over  x''+a x'+b x}=K\eqno(2.13)$$
for the linear equation, and
$$x''x'+2ax'^2+3bx'x+(2ab-b')x^2=M\eqno(2.14)$$
for the nonlinear one, with $b=a^2-a'/2$ and $a$ satisfying the equation
$$a'''=6a''a+7a'^2-16a'a^2+4a^4\eqno(2.15)$$
which is equation XII in the  Chazy classification [14]. Its general
solution can be obtained by putting  $a=-u'/2u$.
Equation (2.15) reduces to
$$u^{(IV)}u-u'''u'+u''^2/2=0\eqno(2.16)$$
 which implies $u^{(V)}$=0, so $u$ is a quartic polynomial in the
independent variable $t$ with one constraint on
its coefficients, because of (2.16).
Given   $a$ and the corresponding $b$, equation (2.14) is integrable by
linearisation
through (2.13). On the other hand, (2.14) violates the Painlev\'e property.
Solving it for $x''$, we find a term
proportional to $x^2/x'$ (or, for that matter, to $1/x'$) which is
incompatible with it.

More cases like the one above could have been derived but this is not
necessary in order to prove our point.
Integrability through linearisation does not require the Painlev\'e
property. On the other hand we do not know of any
systematic way to detect linearisability for a given differential system.
\bigskip
{\scap 3. Discrete integrable systems}
\smallskip
In the case of discrete systems, a difficulty appears from the outset in
the sense that the discrete analogue of
the Painlev\'e property is not well established. One of the properties that
characterises discrete
integrable systems is that of singularity confinement [7].
While analysing a host of integrable mappings it was observed that whenever
a singularity appeared at some
iteration, due to the particular initial conditions,  it disappeared after
some further iterations.
Thus, confinement would have been an excellent candidate for the role of
the discrete analogue of
the Painlev\'e property were it not for the fact that it is not sufficient.
There exist mappings which have
only confined singularities and which are {\sl not} integrable [8]. Another
property which has been proposed as an
indicator of integrability in (rational) mappings is that of the degree
growth of the iterates [15].

Let us illustrate what we mean by degree in a specific example. We consider
a three-point mapping of the form
$\xup=f(x,\xdo;n)$ where $f$ is rational in $x,\xdo$. (The `bar' notation,
which will be used throughout this
section, is a shorthand for the up- and down-shifts in $n$ i.e. $\xup\equiv
x(n+1)$, $x\equiv x(n)$, $\xdo\equiv
x(n-1)$). Starting from some initial conditions $x_0,x_1$ we introduce
homogeneous variables through $x_0=p$,
$x_1=q/r$ and compute the homogeneity degree of the iterates of the mapping
in $q,r$, to which we assign the
same degree 1, while $p$ is assigned the degree 0. Other choices do exist
but the result does not depend on the
particular choice. While the degrees obtained do depend on it, the {\sl
growth} of the degree does not.
Thus for a generic, nonintegrable, mapping the degree growth of the
iterates is exponential [16,17]. On the contrary, for integrable mappings,
the growth is just polynomial. Moreover, a
detailed analysis of discrete Painlev\'e equations [18] and linearisable
mappings [19] has shown that the latter have
even slower growth properties  (which can be used not only as a detector of
integrability but as an indicator of the
integration method). In what follows, we shall examine the results of the
application of the two methods to
integrable discrete systems.

The first case we shall analyse are projective mappings [11]. In perfect
analogy to the continuous case one can
introduce the discrete projective Riccati equations. The starting point is
a linear system for ($N$+1) variables:
$$\Xup_\mu=\sum_{\nu=0}^N A_{\mu\nu}X_\nu \qquad\mu=0,1,\dots,N \eqno(3.1)$$
Introducing again $x_\mu=X_\mu/X_0$, we obtain:
$$\xup_\mu={A_{\mu 0}+\sum_{\nu=1}^N A_{\mu\nu}x_\nu\over A_{0
0}+\sum_{\nu=1}^N A_{0\nu}x_\nu}
\qquad\mu=1,\dots,N
\eqno(3.2)$$
 In fact we have shown [12] that this system can always be rewritten as a
$N$+1-point mapping in
terms of a single object. Clearly the case
$N$=1 is just a homographic (discrete Riccati) mapping for
$x_1$. For $N$=2 we finally find [20,21]:
$$\overline w=\alpha+{\beta \over w}+{1\over w\underline w}
\eqno(3.3)$$
for a quantity $w$ which is obtained from $x_1$, say, through some
homography and $\alpha$, $\beta $ are given in
tems of the $A_{\mu\nu}$.
 Because of the underlying linearisation, any singularity appearing in the
projective
Riccati system is confined in one step. Moreover, the study of the degree
of the iterates [19] shows that there is no
growth at all: the degree is constant. Thus both criteria are satisfied in
this case.

We turn now to the more interesting case of the Gambier mapping [22]. The
latter is, in perfect analogy to the
continuous case, a system of two (discrete) Riccati equations in cascade:
$$\yup={\alpha y+\beta\over \gamma y+\delta} \eqno(3.4a)$$
$$\xup= {a yx+bx+cy+d\over fyx+gx+hy+k} \eqno(3.4b)$$
 where $\alpha$, \dots, $\delta$ and $a$, \dots, $k$ are all functions of
the independent discrete variable $n$.
In [22] it was shown that system (3.4) is not confining unless the
coefficients entering in the equation satisfy certain
conditions. On the other hand the same argument presented in the continuous
case can be transposed here: the
integration of the two Riccati equations in cascade can always be
performed, through reduction to linear second
order mappings. The study  of the degree growth of the iterates of (3.4)
was performed in [19] where it was found
that the growth is always linear, {\sl independently} of the
condition we referred to above.

This result leads naturally to the following generalisation of the discrete
Gambier system, the singularities of
which are, in general, not confined:
$$\yup={\alpha y+\beta\over \gamma y+\delta} \eqno(3.5a)$$
$$\xup= {a( y)x+b( y)\over c( y)x+d( y)} \eqno(3.5b)$$
where $a$, \dots, $d$ are polynomials in $y$ the coefficients of which may
depend on the independent
variable $n$. The study of the degree growth of the iterates of (3.5) is
straightforward. We find that the degree
growth of $x$ is linear. Again, system (3.5) can be integrated in cascade.
On the other hand, (3.5) cannot be written as
a three-point mapping for $x$. Indeed, if we eliminate $y$, $\yup$ between
(3.5a) (3.5b) and the upshift of the latter,
we obtain an equation relating $x$, $\xup$ and $\overline{\xup}$ which is
polynomial in all three variables,
generically {\sl not} linear in $\overline{\xup}$. This does not define a
mapping but rather a correspondence which in
general leads to exponential proliferation of the number of images and
preimages. This correspondence is not
integrable but this is not in contradiction with the integrability of
(3.5). The two systems are not equivalent.

An $(N+1)$-variables extension of the Gambier system can be easily
produced. We have:
$$\xup_0= {\alpha x_0+\beta\over \gamma x_0+\delta} \eqno(3.6)$$
$$\xup_\mu= {a_\mu( x_0,\dots,x_{\mu-1})x_\mu+b_\mu(
x_0,\dots,x_{\mu-1})\over c_\mu(
x_0,\dots,x_{\mu-1})x_\mu+d_\mu( x_0,\dots,x_{\mu-1})}\qquad\mu=1,\dots, N.$$
Again, the degree growth of (3.6)  can be computed leading to a linear
growth and, once more, the singularities of
(3.6) do not confine in general.

Thus, several linearisable systems can be found for which the singularity
confinement gives more restricted
predictions than the degree growth.  We shall comment on this point in the
conclusion.

A last point concerns the discrete analogues of the linearisable systems we
have presented at the end of
section 2. The procedure can be transposed to a discrete setting in a
pretty straightforward way. We have a linear
equation
$${\alpha \xup+\beta+\gamma\xdo+\delta\over \epsilon
\xup+\zeta+\eta\xdo+\theta}= K\eqno(3.7)$$
where $\alpha$,\dots,$\theta$ are all functions of $n$ with  $K$  a
constant, and a nonlinear mapping
$$f(\xdo,x,\xup;n)= M\eqno(3.8)$$
where $f$ is globally polynomial of degree two in all the $x$'s but not
more than linear separately in each of
$\xdo$ and $\xup$. Writing that the l.h.s. of
(3.7) is the same as that of its upshift we get an equation relating
$\xdo$, $x$, $\xup$ and $\overline {\xup}$.
For appropriate choices of $\alpha$,\dots,$\theta$ this four point equation
can be identical (up to unimportant
factors) to the four-point equation obtained from (3.8)
 by writing $f(\xdo,x,\xup;n)=f(x,\xup,\overline {\xup};n$+1). The integration
method is quite similar to that described in the continuous case.
Given  $M$, and starting with $\xdo,x$ at some $n$, one gets $\xup$ from
(3.8). Implementing (3.7) this fixes the
value of $K$. From then on, one integrates the linear equation (3.7) for
all $n$. Since the four-point equation is
always satisfied, this means that $f$ computed at any $n$ has a constant
value, which is just $M$, so
(3.8) is satisfied.

 Several mappings derived in [23] as special limits of
discrete Painlev\'e equations can be linearised in this way.
For instance the nonlinear equation:
$$\left({\xup+x-a\over \zup}-{x\over \zeta }\right)\left({\xdo+x-a\over z}-
{x\over \zeta }\right)-{x^2\over \zeta^2}=M\eqno(3.9)$$
with $a$ a constant, where $z$ and $\zeta$ are defined from a single
arbitrary function $g$ of $n$ through $z=\overline
g+\underline g$, $\zeta=\overline g+ g$, can be solved through the linear
equation:
$${A\xup +B (x-a) +\overline A \xdo\over z\xup+(\zup+z)(x-a)+\zup\xdo
}=K\eqno(3.10) $$
where $A=g^2(\overline g+\underline g)$ and $B=-(\overline g+
g)\overline{\overline g}\underline g-
(\overline{\overline g}+\underline g)\overline g g$. Mapping (3.9) is
generically non-confining unless $g$ is a constant.
\bigskip
{\scap Conclusion}
\smallskip
In this paper we have adressed the question of integrability which does not
necessitate the Painlev\'e
property. We have found that for a large class of integrable, {\sl
linearisable} systems, the Painlev\'e
property is not a prerequisite for integrability. The second-order Gambier
system is the prototype of such a
linearisable equation. Once we find that it can be linearised in the
absence of the Painlev\'e property, it is
straightforward to generalise the Gambier system and to extend it to
$N$ variables (violating the Painlev\'e property but preserving
integrability through linearisability).

Having dispensed of the Painlev\'e property, it is possible to propose a
new method of
linearisation where the derivative of a nonlinear system coincides with
that of a linear one.  The
usefulness of this method has been  illustrated through the derivation of a
linearisable system
which does not satisfy the  Painlev\'e criterion.

At this point, we must stress that the Painlev\'e property can still be
considered as a necessary
condition for integrability provided we qualify the latter. The
integrability with which the  Painlev\'e
property  is associated, often referred to as algebraic integrability,
corresponds to the integration
through IST methods. This is for instance the case of the transcendental
Painlev\'e equations (or most
of the integrable partial differential equations). For these cases, the
Painlev\'e property is necessary
 and we believe, sufficient.
What this paper shows is that for the simpler case of linearisability, the
Painlev\'e property  is superfluous.

In the case of discrete systems the situation is more complicated. It would
appear that what would
play the role of the  Painlev\'e property  is singularity confinement. (The
caveat is that the latter was shown not to be a sufficient condition).
Again, it turned out that
singularity confinement is necessary for integrability through IST methods,
as for instance in the integration
of the  discrete Painlev\'e equations through isomonodromy techniques.
However for integrability through
linearisation, singularity confinement is too restrictive just like the
Painlev\'e property. The study of the
degree growth, on the other hand, shows that this criterion is more
suitable for the detection of integrability
in a larger sense: it identifies all linearisable systems as integrable
with no restrictions whatsoever.
This is at variance with the continuous case where no linearisability
criterion seems to exist (at least none has been
found to date). Moreover, the detailed information on the degree growth is
a most useful
indication of the precise integration procedure.
Thus, although it is not clear whether  the degree growth is the discrete
equivalent to the Painlev\'e property it
can be a most reliable integrability detector.

\bigskip
{\scap Acknowledgments}
\smallskip
S. Tremblay acknowledges a scholarship from le Centre de Coop\'eration
Interuniversitaire Franco-Qu\'eb\'ecois.

\bigskip
{\scap References}
\medskip
\item{[1]} P. Painlev\'e, Acta Math. 25 (1902) 1.
\item{[2]}	E.L. Ince, {\sl Ordinary Differential Equations}, Dover,
London, (1956).
\item{[3]} M.J. Ablowitz and H. Segur, Phys. Rev. Lett. 38 (1977) 1103.
\item{[4]} M. Adler and P. van Moerbeke, {\sl Algebraic completely
integrable systems: a systematic approach},
Perspectives in Mathematics, Academic Press, New York (1988).
\item{[5]} M.J. Ablowitz and P.A. Clarkson, Solitons and the Inverse
Scattering Transform, SIAM, Philadelphia (1981).
\item{[6]} M.D. Kruskal, A. Ramani and B. Grammaticos, NATO ASI Series C
310, Kluwer (1989) 321.
\item{[7]} B. Grammaticos, A. Ramani and V. Papageorgiou, Phys. Rev. Lett.
67 (1991) 1825.
\item{[8]} J. Hietarinta and C. Viallet, Phys. Rev. Lett. 81, (1998) 325.
\item{[9]} A. Ramani, B. Dorizzi, B. Grammaticos, Phys. Rev. Lett. 49
(1982) 1539.
\item{[10]} B. Dorizzi, B. Grammaticos, A. Ramani, J. Math. Phys. 25 (1984)
481.
\item{[11]} B. Grammaticos, A. Ramani, P. Winternitz, Phys. Lett. A 245
(1998) 382.
\item{[12]} S. Lafortune, B. Grammaticos, A. Ramani, Physica A 268 (1999) 129.
\item{[13]} B. Gambier, Acta Math. 33 (1909) 1.
\item{[14]} J. Chazy, Acta Math. 34 (1911) 317.
\item{[15]} M.P. Bellon and C.-M. Viallet, Comm. Math. Phys. 204 (1999) 425.
\item{[16]} V.I. Arnold, Bol. Soc. Bras. Mat. 21 (1990) 1.
\item{[17]} A.P. Veselov, Comm. Math. Phys. 145 (1992) 181.
\item{[18]} Y. Ohta, K.M. Tamizhmani, B. Grammaticos and A. Ramani, Phys.
Lett. A. 262 (1999) 152.
\item{[19]} A. Ramani, B. Grammaticos, S. Lafortune and Y. Ohta, {\sl
Linearisable mappings and the low-growth
criterion}, preprint (1999).
\item{[20]} A. Ramani, B. Grammaticos, G. Karra, Physica A 180 (1992) 115.
\item{[21]} A. Ramani, B. Grammaticos, K.M. Tamizhmani, S. Lafortune,
Physica A 252 (1998) 138.
\item{[22]} B. Grammaticos, A. Ramani, Physica A 223 (1996) 125.
\item{[23]} A. Ramani, B. Grammaticos, Y. Ohta, B. Grammaticos, {\sl
Discrete integrable systems from continuous
Painlev\'e equations through limiting procedures}, preprint (1999).

\end